\input{aipcheck}

\documentclass[
    ,final            
  ]
  {aipproc}

\layoutstyle{6x9}
\usepackage{physics,myphysics}

\begin{document}

\def \nngg {$\nu\bar\nu\gamma(\gamma)$}
\def\GeV{\ifmmode {\mathrm{\ Ge\kern -0.1em V}}\else
                   \textrm{Ge\kern -0.1em V}\fi}%
\def\eV{\ifmmode {\mathrm{\ e\kern -0.1em V}}\else
                   \textrm{e\kern -0.1em V}\fi}%
\def\TeV{\ifmmode {\mathrm{\ Te\kern -0.1em V}}\else
                   \textrm{Te\kern -0.1em V}\fi}%
\def\gravin{\ensuremath{\tilde\mathrm{G}}}%
\def\DELTAM{\ensuremath{\Delta m}}%
\def\Msel{\ensuremath{m_{\tilde\e_\mathrm{L}}}}%
\def\Mse{\ensuremath{m_{\tilde\mathrm{e}}}}%
\def\Mselr{\ensuremath{m_{\tilde\e_\mathrm{L,R}}}}%
\def\Mchi{\ensuremath{m_{\tilde\chi^0_1}}}%
\def\Mchii{\ensuremath{m_{\tilde\chi^0_2}}}%
\def\Mcha{\ensuremath{m_{\tilde\chi^\pm_1}}}%
\def\Msnu{\ensuremath{m_{\tilde\nu}}}%
\def\Msml{\ensuremath{m_{\tilde\mu_\mathrm{L}}}}%
\def\Mstat{\ensuremath{m_{\tilde\tau_2}}}%
\def\MP{\ensuremath{m_{\mathrm{P}}}}%
\def\MG{\ensuremath{m_{\gravin}}}%
\def\susyl#1{\ensuremath{\tilde{#1}_\mathrm{L}}}%
\def\susyr#1{\ensuremath{\tilde{#1}_\mathrm{R}}}%
\def\susylr#1{\ensuremath{\tilde{#1}_\mathrm{L,R}}}%
\def\tanb{\ensuremath{\tan \beta}}%
\def\EE{\ensuremath{\mathrm{e^+e^-}}}

\title{Searches for Gauge-Mediated SUSY Breaking Topologies with the L3
Detector at LEP}


\author{M.~Gataullin}{
  address={Department of Physics, California Institute of Technology, 
MC 256-48, Pasadena, CA 91125, USA}
}

\author{S.~Rosier}{
  address={LAPP Annecy-le-Vieux,  Chemin du Bellevue, 
BP 110 F-74941, CEDEX France}
}

\author{L.~Xia}{
  address={Argonne National Laboratory, Argonne, IL 60439, USA}
}

\author{H.~Yang}{
  address={Department of Physics, University of Michigan,
 Ann Arbor, MI 48109,  USA}
}

\begin{abstract}
Searches for topologies predicted by gauge-mediated SUSY breaking
models were performed using data collected with the L3 detector at
LEP. All possible lifetimes of the 
next-to-lightest SUSY particle (NLSP), 
neutralino or scalar tau, were considered. 
No evidence for these new phenomena was found and limits on the
production cross sections and sparticle masses were derived. 
A scan over the parameters of the  minimal GMSB model was performed,
leading to lower limits of 62.2~\GeV, 11~\TeV, and 0.07~\eV\
on the NLSP mass, the mass scale parameter $\Lambda$, and the
gravitino mass, respectively. The status of the LEP combined
searches is also discussed.
\end{abstract}

\maketitle


\section{Introduction}

In this paper, we briefly review results of the searches for 
manifestations of the gauge-mediate Supersymmetry breaking models
(GMSB) in $\EE$ collisions at LEP. In GMSB, the lightest SUSY particle
is always the gravitino, with a mass in the range of $10^{-2}-10^4~\eV$,
whereas the next-to-lightest SUSY particle (NLSP) is typically 
the scalar tau or the lightest neutralino.
The minimal GMSB model can be described by a set of
the following parameters: $\MG$, the gravitino mass; $\Lambda$, 
the universal mass scale of SUSY particles; $M_m$, the messenger mass;
$N_m$, the messenger index; ~\tanb,
the ratio between the vacuum expectation values of the two Higgs doublets;
and sign$(\mu)$, the Higgs mixing parameter. 
Because  the lifetime of the NLSP
depends on the gravitino mass, NLSP
decays both inside and outside the 
detector volume had to be considered.

Data collected by the L3 detector at LEP in the
years from 1998 through 2000 were used. They correspond to an
integrated luminosity of 619~\pb\ at center-of-mass energies
$\sqrt{s}= 189 -209\GeV$. All results given below are considered
to be preliminary. All limits are derived at the 95\% confidence level.

\section{The \chinon\ NLSP Scenario}

In the case of the neutralino NLSP, the GMSB signature
is the pair-production of neutralinos, each decaying into a photon and a 
gravitino:  $\mathrm{e}^+\mathrm{e}^- \rightarrow \tilde\chi^0_1 
\tilde\chi^0_1 \rightarrow
\tilde\mathrm{G} \gamma \tilde\mathrm{G} \gamma$.

If both neutralinos decay promptly, 
the experimental signature of this process is very clean, involving
 events with two energetic acoplanar photons. 
Searches for this signature were motivated by 
the rare ee$\gamma\gamma$ event observed by the CDF experiment.
No evidence for anomalous production of such events 
was found and the GMSB interpretation
of the CDF event was excluded~\cite{bib:l3sphot}.

\begin{figure}
  \resizebox{19pc}{!}{\includegraphics{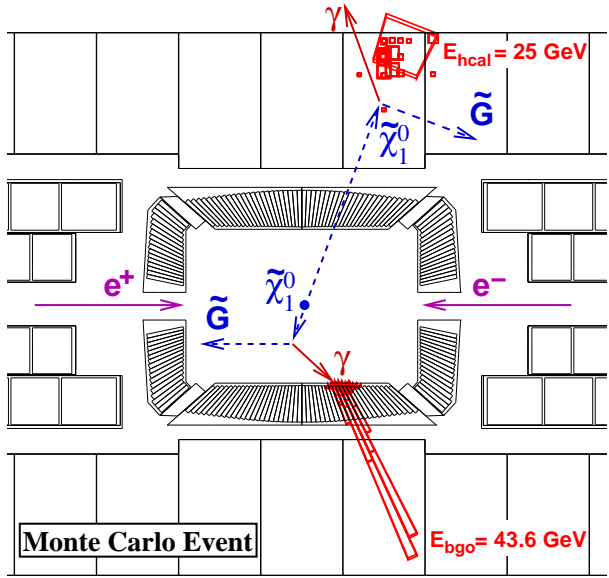}}
\caption{
 A simulated $\EE\rightarrow
\chinon\chinon\,\rightarrow \gravin\gravin\gam\gam$ event with two 
non-pointing photons,  where $\Mchi=95~\GeV$ and $\sqrt{s}=207~\GeV$.
 In this event one of the neutralinos decayed in front
of the electromagnetic calorimeter, 
while the other decayed and produced a shower inside the hadron calorimeter.
}
\label{fig:nonpevent}
\end{figure}

However, for $\MG \sim 100~\eV$, the proper decay length of the
 neutralino,
$c\tau_{\chinon}$, can become comparable to or even larger than 
the size of the L3 detector.
In this scenario, one or both of the produced neutralinos 
may decay 
 within the sensitive volume of the
 detector, but at a distance from the primary vertex.
This would lead to events  with non-pointing photons, such as shown 
in Figure~\ref{fig:nonpevent}.

To improve the sensitivity of this search, non-pointing photons
were identified using not only the electromagnetic (ECAL) but also 
the hadron calorimeter (HCAL) of L3. In the ECAL,
such photons were selected by studying  the transverse profile of 
electromagnetic showers. For the signatures with neutralino
decays in the HCAL, the Standard 
Model background was absent. However, the HCAL response to 
electromagnetic showers had to be measured {\it in situ} at LEP.
A dedicated study was performed 
using events from the Bhabha scattering process, $\EE\rightarrow\EE$, 
in which both scattered electrons passed through the
gaps between the ECAL barrel and endcaps and produced two showers 
in the HCAL. In addition, special veto cuts were 
applied to reject cosmic ray backgrounds.

\begin{figure}
\begin{tabular}{cc} 
  \resizebox{17pc}{!}{\includegraphics{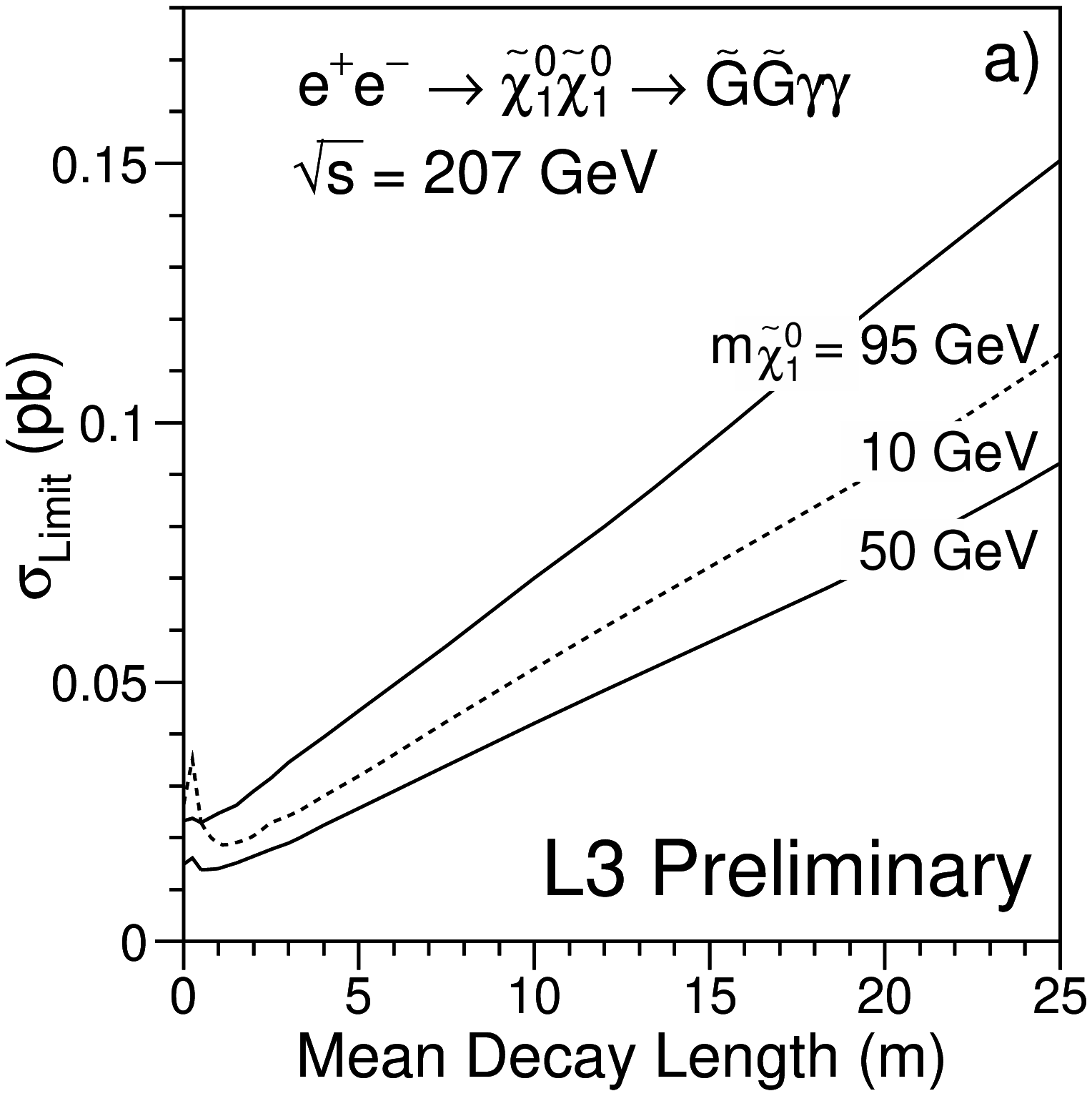}} &
  \resizebox{17pc}{!}{\includegraphics{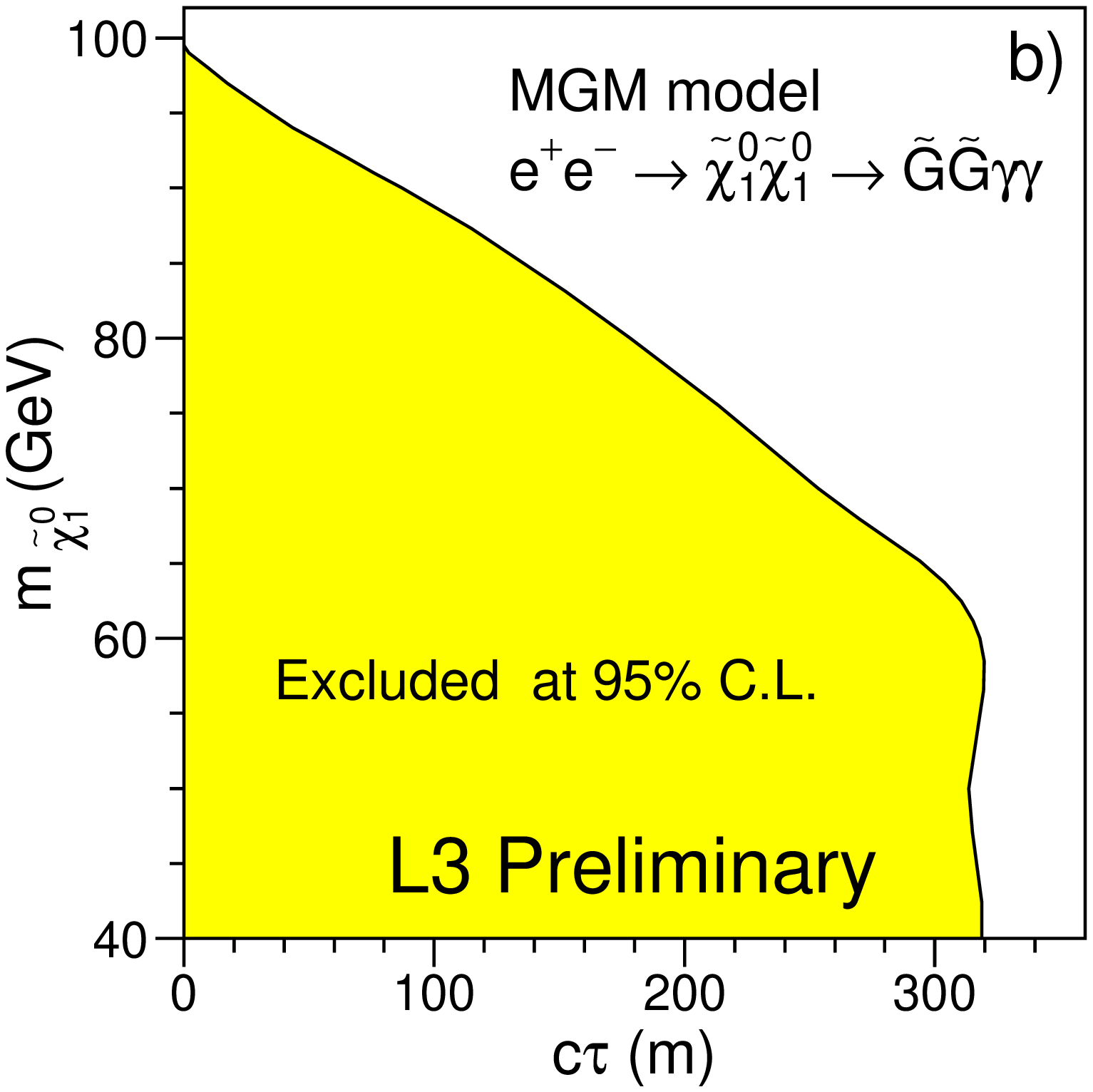}} \\
  \resizebox{15pc}{!}{\includegraphics{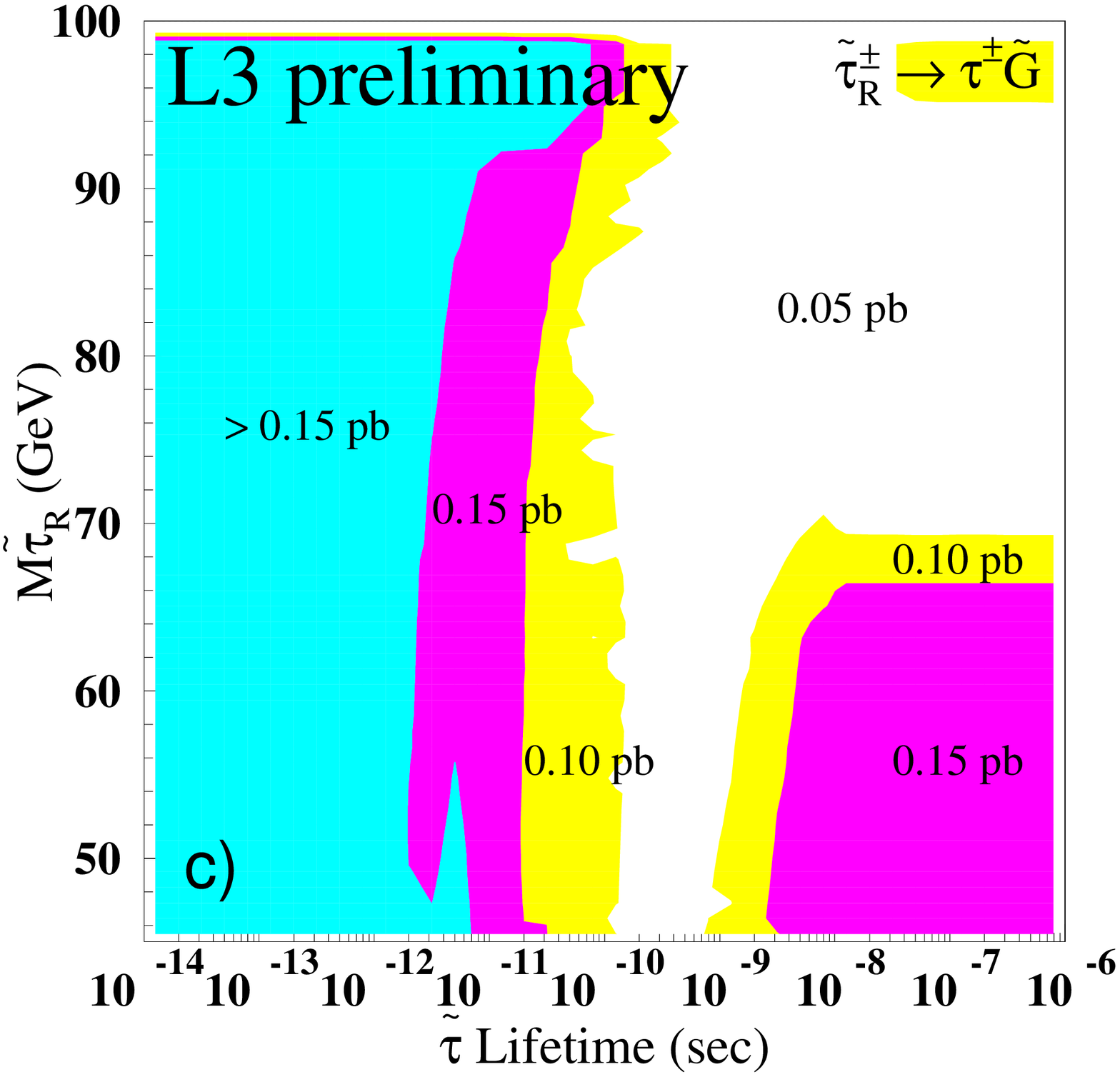}} &
  \resizebox{14.7pc}{!}{\includegraphics{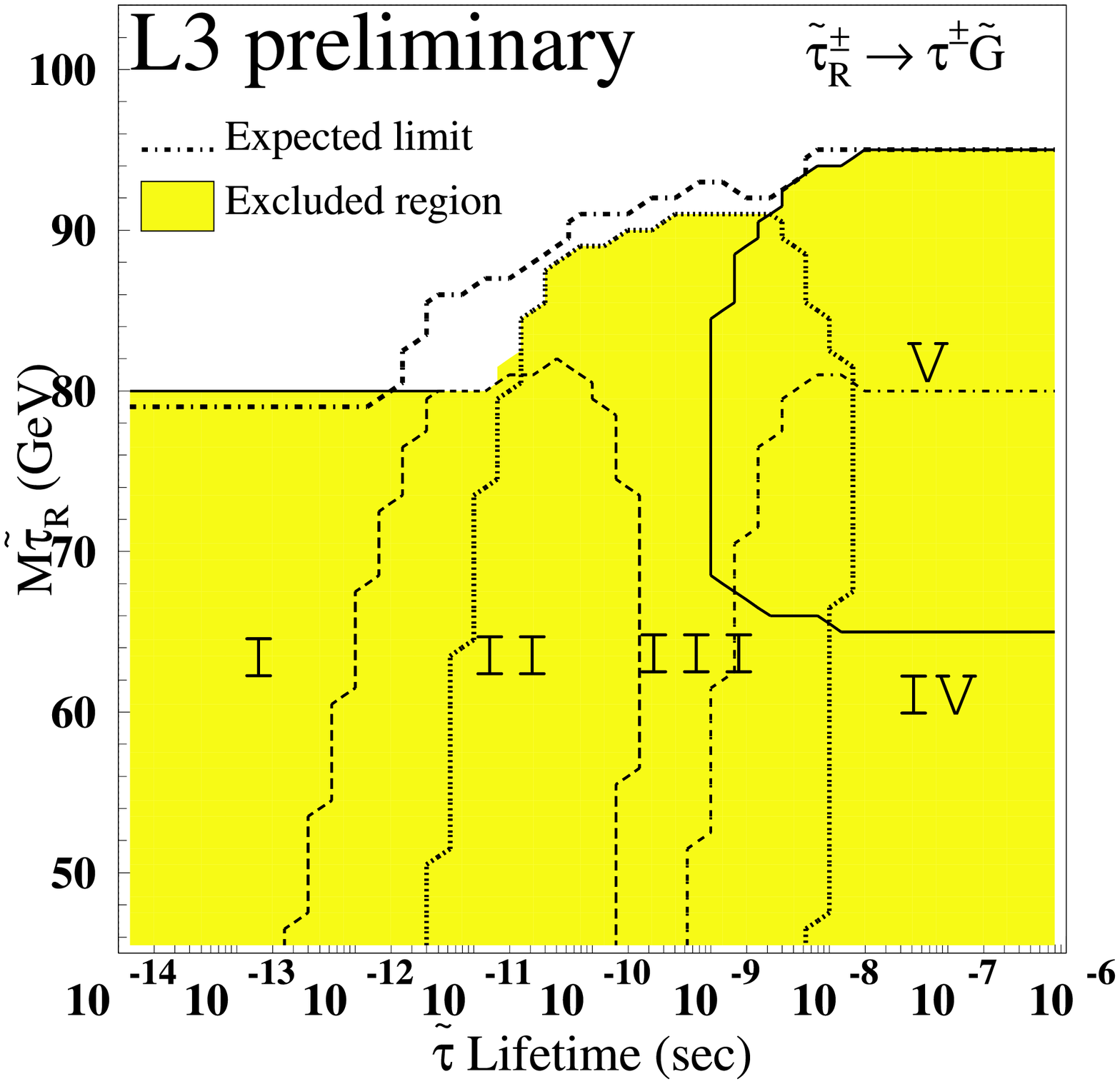}}
\begin{minipage}{.03\linewidth}
\vspace*{-2.1cm} 
\hspace*{-5.35cm} 
        {\sffamily  d)}
    \end{minipage}
\end{tabular}
\caption{
{\bf a)}~Upper 
limits on the $\EE\rightarrow\chinon\chinon$
 production cross section for three 
different neutralino mass hypotheses $\Mchi=95,50,\,\mathrm{and}\,10~\GeV$,
as functions of the mean $\chinon$ decay length in the laboratory frame:
$L_{\chinon}\,=\,\gamma\beta c\tau_{\chinon} \, $.
{\bf b)}~Region excluded in the  $(c\tau_{\chinon},\Mchi)$ plane 
under the assumptions of the MGM model~\cite{bib:mgm}.
{\bf c)}~Upper limit on the 
$\EE\rightarrow\tilde{\tau}\tilde{\tau}$ production cross section 
at $\sqrt{s} = 205~\GeV$, as a function of the $\tilde{\tau}$ mass and 
lifetime.
{\bf d)}~Region 
excluded in the  $(\tau_{\tilde{\tau}},m_{\tilde{\tau}})$ plane
using five sets of different selection cuts: 
I prompt decay; II and III intermediate
decay lengths; IV and V long-lived $\tilde{\tau}$.
}
\label{fig:nonplims}
\end{figure}

No events with non-pointing photons 
were found and limits on the $\EE\rightarrow
\chinon\chinon$ production cross section were derived 
(see Figure~\ref{fig:nonplims}a).
These limits were then translated into an
excluded region in the $(c\tau_{\chinon},\Mchi)$ plane.
As shown in Figure~\ref{fig:nonplims}b,  neutralino masses 
$ \Mchi < 88\GeV$ 
were excluded for $c\tau_{\chinon}$ values smaller than 100~m.
A more detailed description of this search
 can be found in Reference~\cite{bib:my_thesis}.

\section{The $\tilde{\tau}$ NLSP Scenario}
In the case of the scalar tau NLSP, the pair production  process
$\EE\rightarrow\tilde{\tau}\tilde{\tau}\rightarrow\tau\gravin\tau\gravin$
was expected to be the main experimental signature. Since the corresponding
final state topology depended strongly on the scalar tau lifetime, 
$\tau_{\tilde{\tau}}$, five
different event selections were developed.

 A search for acoplanar leptons
covered the case of prompt $\tilde{\tau}$ decays. 
For this signature, the selection cuts were
the same as in the search for $\tilde{\tau}$ production in gravity
mediated Supersymmetry~\cite{bib:l3sleptons}. 
In the case of intermediate decay lengths, a
search for charged tracks with large impact parameters and kinks
was performed. Finally, long-lived scalar taus  were 
searched for by measuring the dE/dx parameter of charged tracks. 

The obtained limits on the $\EE\rightarrow\tilde{\tau}\tilde{\tau}$
cross section are shown in Figure~\ref{fig:nonplims}c. These limits
were then translated into a limit on the $\tilde{\tau}$ mass:
$m_{\tilde{\tau}} > 80~\GeV$ (see Figure~\ref{fig:nonplims}d). 
A detailed description of this search can be found in 
Reference~\cite{bib:xia_thesis}.

\section{Combinations of Search Results}

To combine results of the searches performed for all possible
NLSP scenarios, including those not described above,
a scan over the GMSB parameter space was performed~\cite{bib:gmsbnote}.
In total, $5.6\cdot 10^6$ points in the parameter space were tested.
At each point the complete mass spectrum, production cross sections, and
branching ratios were calculated.
A point in the parameter space was considered excluded 
if it was kinematically accessible and the calculated
 cross section was higher
than the cross section limit for the corresponding discovery channel.
The results of this scan are summarized in Table~\ref{tab:scan}.

For the $\tilde{\tau}$ NLSP scenario,
a combination of the searches 
by all four LEP experiments was performed, excluding
$\tilde{\tau}$ masses below 86.9~\GeV~\cite{bib:lepsusy1}.
A LEP-wide combination also exists for the  short-lived \chinon\ 
NLSP scenario~\cite{bib:lepsusy2}.

\begin{table}
\begin{tabular}{llllll} 
\hline
 \tablehead{1}{l}{b}{Messenger Index}     &
 \tablehead{1}{l}{b}{1}     &
 \tablehead{1}{l}{b}{2}     &
 \tablehead{1}{l}{b}{3}     &
 \tablehead{1}{l}{b}{4}     &
 \tablehead{1}{l}{b}{5}         \\ \hline
  $\Lambda$~(\TeV)            & 47.3  & 25.5  & 16.8   & 13.0 & 11.0 \\ 
  $M_{mess}$~(\TeV)     & 56    & 32    & 24 & 24 & 18 \\
  $\MG$~(\eV)        & 0.63   & 0.19 & 0.11 & 0.10 & 0.07 \\ 
  $\Mchi$~(\GeV)     & 62.2 & 65.9  & 65.6  & 65.2& 72.2 \\ 
  $m_{\tilde{\tau}_1}$~(\GeV)    & 62.3 & 62.2 & 62.3 & 62.7 & 62.8 \\ 
  $m_{\tilde\e_\mathrm{R}}, m_{\tilde\mu_\mathrm{R}}$~(\GeV) 
                                    & 89.6 & 73.6 & 71.2 & 66.9 & 67.1 \\
\hline
\end{tabular}
\caption{Lower limits on the parameters of the  minimal GMSB 
and sparticle masses for all possible NLSP scenarios.
} 
\label{tab:scan} 
\end{table}

\begin{theacknowledgments}
We would like to thank all our colleagues from the L3 collaboration.
This work was supported in part by  the 
 U.S. Department of Energy Grant No.~DE-FG03-92-ER40701.
\end{theacknowledgments}



\bibliographystyle{aipproc}   

\end{document}